\documentclass[manuscript]{aastex}
\shorttitle{hard electron spectrum in the afterglow of GRB 091127}
\shortauthors{Zhang et al.}

\begin{document}
\title{Modelling the Multi-band Afterglow of GRB 091127: Evidence of a Hard Electron Energy Spectrum with an Injection Break}
\author{Qiang Zhang\altaffilmark{1,2}, Yong-Feng Huang\altaffilmark{3,4}, Hong-Shi Zong\altaffilmark{1,2,5}}
\altaffiltext{1}{Department of Physics, Nanjing University, Nanjing 210093, China; zonghs@nju.edu.cn}
\altaffiltext{2}{Joint Center for Particle, Nuclear Physics and Cosmology, Nanjing 210093, China}
\altaffiltext{3}{School of Astronomy and Space Science, Nanjing University, Nanjing 210093, China; hyf@nju.edu.cn }
\altaffiltext{4}{Key laboratory of Modern Astronomy and Astrophysics (Nanjing University), Ministry of
Education, Nanjing 210093, China}
\altaffiltext{5}{State Key Laboratory of Theoretical Physics, Institute of Theoretical Physics, CAS, Beijing, 100190, China}

\begin{abstract}
 The afterglow of GRBs is believed to originate from the synchrotron emission of shock-accelerated electrons produced by the interaction between the outflow and the external medium. The accelerated electrons are usually assumed to follow a power law energy distribution with an index of $p$. Observationally, although most GRB afterglows have a $p$ larger than 2, there are still a few GRBs suggestive of a hard ($p<2$) electron spectrum. The well-sampled broad-band afterglow data for GRB 091127 show evidence of a hard electron spectrum and strong spectral evolution, with a spectral break moving from high to lower energies. The spectral break evolves very fast and cannot be explained by the cooling break in the standard afterglow model, unless evolving microphysical parameters are assumed. Besides, the multi-band afterglow light curves show an achromatic break at around 33 ks. Based on the model of a hard electron spectrum with an injection break, we interpret the observed spectral break as the synchrotron frequency corresponding to the injection break, and the achromatic break as a jet break caused by the jet-edge effect. It is shown that the spectral evolution and the multi-band afterglow light curves of GRB 091127 can be well reproduced by this model.
\end{abstract}


\section{INTRODUCTION}
Gamma-Ray bursts (GRBs) are the most energetic stellar explosions in the universe; they produce a short prompt $\gamma$-ray emission followed by a multi-band afterglow that can be observed up to several years. The afterglow of GRBs is believed to originate from the synchrotron emission of shock-accelerated electrons produced by the interaction between the outflow and the external medium \citep{Rees92,Mes93,Mes97,Sari98,Cheva2000}. When the blast wave enters a self-similar phase described by the Blandford-McKee self-similar solution \citep{BM76}, the hydrodynamics is mainly determined by six free parameters, i.e. the total energy of the blast wave $E$, the jet half-opening angle $\theta_{\rm{j}}$, the fractions of shock energy carried by electrons and magnetic fields ($\epsilon_{\rm{e}}$ and $\epsilon_{\rm{B}}$), the ambient density $n$ and the electron spectral index $p$. The temporal and spectral indices of the afterglow emission is largely determined by the electron spectral index $p$, which is dependent only on the underlying micro-physics of the acceleration process.

Particle acceleration is usually attributed to the Fermi process \citep{Fermi54}, in which particles bounce back and forth across the shock to gain energy. Particles accelerated by this mechanism follow a power-law energy distribution $N\left(E\right){\rm{d}}E \propto E^{-p}{\rm{d}}E$, with a cut-off at high energies. Some analytical and numerical studies indicate a nearly universal spectral index of $p\sim 2.2-2.4$ \citep{Kirk00,Acht01,Bed98,Lem03,Spit08}, though other studies suggest that there is a large range of possible values for $p$ of $1.5-4$ \citep{Bar04}. Observationally, the value of $p$ can be estimated from the spectral analysis of the multi-band afterglow \citep{Cheva2000,Panai02,Starl08,Curran09} or the X-ray data alone \citep{Shen06,Curran10}. Both studies, however, show that the observed values of $p$ are inconsistent with a single universal value, but show a rather wide distribution.

Moreover, observations of some GRB afterglows suggest a hard electron spectrum with an index $p<2$ \citep{Panai01a}. GRB 010222 was one of the first afterglows
seen with such a hard electron spectrum \citep{Mas01,Stan01}, which motivated theoretical studies in that direction \citep{B01,Dai2001,Huang06,Resmi08}. Different hard-spectrum models were assumed to explain the afterglow of GRB 010222 \citep{Sag01,Cow01,Panai02,B04,Resmi08}, though other explanations, e.g. continuous energy injection \citep{Bjor02}, could also reproduce the observed evolution of this afterglow. Other GRB afterglows, e.g. GRB 020813 \citep{Cov03,But03}, GRB 041006 \citep{Misra05}, showing similar characteristics could also be explained with a hard electron spectrum \citep{Resmi08}.

 For a hard electron spectrum, a cut-off at the high energy end is required to keep the total energy from diverging. The theories of a hard electron spectrum can be divided into two categories by and large. One kind of models assume that a hard electron energy distribution can extend to a maximum electron Lorentz factor $\gamma_{\rm{M}}$, beyond which there is an exponential cut-off \citep{B01,Dai2001}. The other kind of models assume that a hard electron spectrum terminates at some electron Lorentz factor $\gamma_{\rm{b}}$, above which the electron distribution steepens to another power law with the index $p>2$ \citep{Panai01,B04,Resmi08,Wang12}. In this paper, we call $\gamma_{\rm{b}}$ an ``injection break'' as done by \citet{B04} and \citet{Resmi08}. The evolution of $\gamma_{\rm{M}}$ or $\gamma_{\rm{b}}$ has so far not been well understood, and different expressions were assumed in the literature. We distinguish between these two kinds of models by naming them the ``single power-law hard spectrum (SPLH)'' model and the ``double power-law hard spectrum (DPLH)'' model, respectively.

 Although some GBR afterglows could be well explained by the hard-spectrum model \citep{Resmi08}, we should note that all the afterglows referred above show a shallow-to-steep decay in the optical and/or X-ray light curves, with an initial decay slope $\sim0.5-0.8$ steepening to $\sim1.3-1.4$ at around $0.5$ d, and the optical/X-ray spectral indices are in the range of $\sim 0.6-1.0$\citep{Resmi08}. These characteristics, however, are ubiquitous among the canonical afterglow light curves observed in the era of {\it Swift} \citep{Nou06,Zhang06}, and the shallow decay are usually explained by assuming continuous energy injection to the decelerating blast wave in the case of $p>2$\citep{Dai98,Zhang01,Rees98,Sari00}. Therefore, the explanation with a hard electron spectrum seems to be dubious, and can be confused with the continuous energy injection model especially when the spectral information is missing \citep{Kumar15}.

GRB 091127, at a redshift of $z=0.49$ \citep{Cucch09,Thone09}, has high-quality broad-band afterglow data \citep{Filgas11,Troja12}. These data allow us to test several proposed emission models and outflow characteristics in unprecedented detail. The broad-band spectral energy distribution (SED) of the afterglow shows evidence of a hard electron spectrum and strong spectral evolution, with a break frequency moving from high to lower energies. Based on the SPLH model with a spectral index of $p=1.5$, \citet{Filgas11} interpreted this spectral break as the cooling break in the case of a homogeneous interstellar medium (ISM) circum-burst environment. However, the observed spectral break evolves much faster ($\propto t^{-1.23}$) than the cooling break ($\propto t^{-1/2}$). To solve this problem, some microphysical parameters (e.g. $\epsilon_{\rm{B}}$) were required to evolve with time. Indeed, modifications of the standard afterglow model with evolving microphysical parameters ($\epsilon_{\rm{e}}$ or/and $\epsilon_{\rm{B}}$) have been proposed to explain the X-ray afterglow plateaus \citep{Ioka06}, chromatic light-curve breaks \citep{Panai06}, afterglow rebrightenings \citep{Kong10}, or some other difficulties encountered with observations \citep{Van14}. So far, a complete knowledge of the microphysical processes is still lacking. How a parameter evolves mainly depends on which phenomenon to be explained, and sometimes the evolution of $\epsilon_{\rm{e}}$ and $\epsilon_{\rm{B}}$ would have to conspire to match with certain observations, which makes this scenario seem ad hoc and contrived \citep{Panai06}.

 Based on the DPLH model proposed by \citet{Resmi08}, we show in this paper that the observed spectral break can be well explained by the injection break frequency $\nu_{\rm{b}}$ and the observed spectral evolution is the result of $\nu_{\rm{b}}$ crossing the optical/NIR bands. Previous
studies usually assume $\nu_{\rm{b}}$ to be above the X-ray band even at late times \citep{Panai02,Resmi08}, thus the spectral evolution caused by $\nu_{\rm{b}}$ could not be observed. Therefore, with the high-quality afterglow data of GRB 091127, it may be the first time we see the evolution
of the injection break in a hard electron spectrum.

Our paper is organized as follows. We summarize the observational facts of GRB 091127 in Section \ref{data}. The model of a double power-law electron spectrum
is described in Section \ref{model}. In Section \ref{fitting}, we constrain the free parameters in this model, and fit the multi-band afterglow light curves. Finally, we sum up our results and give a brief discussion in Section \ref{conclusion}. Throughout the paper, the convention $F_{\nu}\propto\nu^{-\beta}t^{-\alpha}$ is followed, and we use the standard notation $Q_x=Q/10^{x}$ with $Q$ being a generic quantity in cgs units. We also assume a concordance cosmology of $H_0=71 ~\rm{km}~ \rm{s}^{-1} \rm{Mpc}^{-1}$, $\Omega_{\rm{M}}=0.27$, $\Omega_\Lambda=0.73$.
All the quoted errors are given at $1\sigma$ confidence level.

\section{OBSERVATIONAL FACTS} \label{data}
GRB 091127 triggered the {\it Swift}/BAT at 23:25:45 UT on 2009 November 27 \citep{Troja09},and was also observed by {\it Konus-Wind}, {\it Suzaku}, and {\it Fermi}/GBM. The measured duration of $T_{90}$ in the $15-350$ keV band is $7.1\pm0.2$ s \citep{Stama09}. The time-averaged spectrum is adequately fit by a Band function \citep{Band93} with $E_{\rm{peak}}=45\pm3$ keV, $\alpha=-1.37\pm0.07$, $\beta=-2.31\pm0.03$, and the total fluence in the $8-1000$ keV energy range is $(1.9\pm0.2)\times10^{-5}$ erg cm$^{-2}$ \citep{Troja12}. With a redshift of $z=0.49$, the isotropic equivalent energy of GRB 091127 is  $E_{\gamma,\rm{iso}}=(1.1\pm0.1)\times 10^{52}$ erg \citep{Troja12}. In addition, GRB 091127 is an event associated with SN 2009nz \citep{Cobb10}.

\subsection{X-ray Afterglow}
Due to an Earth limb constraint, {\it Swift}/XRT started to follow-up GRB 091127 about 53 minutes after the BAT trigger \citep{Evans09b}, and the X-ray observations continued for around 50 days. The X-ray spectra were fitted with an absorbed power-law, with the photon index $\Gamma_{\rm{X}}\sim 1.8$, and the host absorbing column density $N_{\rm{H}}\sim 1.3\times10^{21}$ cm$^{-2}$ \citep{Filgas11}. The X-ray light curve could be described by a smoothly broken power-law model \citep{Beue99}, with an initial decay slope $\alpha_{\rm{X,1}}=1.02\pm0.04$ steepening to $\alpha_{\rm{X,2}}=1.61\pm0.04$ at the break time $t_{\rm{bk}}\approx33$ ks \citep{Filgas11}. Similar results were given by \citet{Troja12}.

\subsection{Optical Afterglow}\label{opt}
 The optical afterglow of GRB 091127 was monitored by {\it Swift}/UVOT, the 2 m Liverpool Telescope (LT), the Faulkes Telescope South (FTS), SkycamZ, and the Gamma-Ray burst Optical Near-infrared Detector (GROND)\citep{Troja12,Filgas11}. LT began observing the burst 141 s after the BAT trigger,
and located the position of the transient at RA(J2000)=$02^h 26^m 19^{s}.89$, and Dec(J2000)=$-18^{\circ}57'08.6''$ \citep[uncertainty of $0.5''$;][]{Smith09}. GROND started observations 58 minutes after the trigger with very high-quality data in the optical/NIR bands \citep{Updike09}. Using the well-sampled NIR/optical data obtained with GROND, \citet{Filgas11} gave a detailed study of the multi-color light curves and the broad-band SEDs from NIR to X-ray. We summarize their main results in the following.

The optical/NIR light curves are also well fitted with a smoothly broken power-law after subtracting the SN-bump at late times. The initial decay slopes are slightly different among the optical bands, ranging from 0.32 to 0.43. The average value is $\alpha_{\rm{opt,1}}=0.38\pm0.04$, it then steepens to $\alpha_{\rm{opt,2}}\sim1.6$ at the break time around 33 ks. The decay of the NIR light curves is slightly shallower but with larger inaccuracies compared with that of optical bands, because the fitting is affected by the somewhat lower signal-to-noise ratio of the NIR data \citep{Filgas11}. The most notable feature is the strong spectral evolution in the optical/NIR bands, the spectral
index $\beta_{\rm{opt/NIR}}$ rises from $\sim0.23$ to $\sim0.80$ between 3 and 300 ks.  The broad-band SEDs before $\sim110$ ks could be fitted with a smoothly broken power-law when $\beta_{\rm{opt/NIR}}=\beta_{\rm{X}}-0.5$ was fixed. The derived high-energy spectral index is $\beta_{\rm{X}}=0.75\pm0.004$. The break frequency decreases from $\sim28.7$ eV to $\sim0.3$ eV, and could be fitted to scale as $\propto t^{-1.23\pm0.06}$. The SEDs at later times are consistent with a simple power-law, and the spectral indices in the optical/NIR bands are in agreement with that of X-ray afterglow within $1\sigma$ errors.
The SEDs require the break frequency to cross the optical bands, which results in the spectral evolution and the very smooth breaks in the optical/NIR light curves. Fits of the optical/NIR data alone as well as the broad-band SEDs imply no host dust extinction.

\citet{Troja12} studied the multi-band afterglow of GRB 091127 using the X-ray data from {\it Swift}/XRT and {\it Chandra} and the optical data from UVOT, LT, FTS, and SkycamZ. The optical light curves show
a gradual shallow-to-steep decay at around 30 ks as well. Their obtained decay slopes are slightly different from those given by \citet{Filgas11}, mainly because they fitted the light curves with a triple broken power-law. They also found a spectral break evolution with a decay index $-1.5\pm 0.5$.  However, the sparse sampling in the optical bands forbade them to derive the detailed evolution of the spectral index and break frequency as done by \citet{Filgas11}.

\section{MODEL} \label{model}
The afterglow of GRB 091127 shows two important properties: the spectral evolution in the optical/NIR bands, and the achromatic break at around 33 ks in both X-ray and optical/NIR light curves. Moreover, a hard electron spectrum is required due to the flat spectrum in the optical/NIR band at early times. Based on the DPLH model and using the assumed evolution function of the injection break proposed by \citet{B01} and \citet{Resmi08}, we interpreted the observed spectral evolution as the injection break frequency $\nu_{\rm{b}}$ crossing the optical/NIR bands, and explain the achromatic break as a jet break due to the jet-edge effect \citep{Mes99} as proposed by \citet{Filgas11} and \citet{Troja12}. Therefore, the multi-band afterglow of GRB 091127 is a joint result of the spectral evolution caused by the injection break and the jet-edge effect. In the following, we give a brief description of the DPLH model and present the useful formula which will be used in Section \ref{fitting} (see \citet{Resmi08} for more details).

The double-slope electron energy distribution with slopes $p_1$ and $p_2$ is represented as \citep{Resmi08,Wang12}
\begin{equation}
N\left(\gamma_{\rm{e}}\right)= C_{\rm{e}}\left\{
\begin{array}{ll}
\left(\frac{\gamma_{\rm{e}}}{\gamma_{\rm{b}}}\right)^{-p_1}, &\gamma_{\rm{m}}<\gamma_{\rm{e}}<\gamma_{\rm{b}},  \\
\left(\frac{\gamma_{\rm{e}}}{\gamma_{\rm{b}}}\right)^{-p_2}, &\gamma_{\rm{b}}<\gamma_{\rm{e}}<\gamma_{\rm{M}},  \\
\end{array}
\right.
\end{equation}
where $C_{\rm{e}}$ is the normalization constant, $\gamma_{\rm{m}}$ and $\gamma_{\rm{M}}$ are minimum and
maximum electron Lorentz factors, respectively, and $\gamma_{\rm{b}}$ is the injection break. The electron spectral indices $1<p_1<2$, $p_2>2$. The definition of $\gamma_{\rm{M}}$ is unimportant, and can be set to infinity \citep{Resmi08}.
The physical origin of $\gamma_{\rm{b}}$ is not clear, but one can simply assume that it is a function of $\gamma$ alone \citep{B01,Resmi08},
\begin{equation}
\gamma_{\rm{b}}=\xi \left(\beta \gamma\right)^q,
\end{equation}
where $\xi$ is a constant of proportionality, $\beta=\sqrt{1-\gamma^{-2}}$ is the dimensionless bulk velocity, $q$ is assumed to be a constant for simplicity.

According to this electron energy distribution and the jump conditions for a relativistic shock, the electron number density and energy density of the shocked medium can be written as two integrals: $\int_{\gamma_{\rm{m}}}^{\gamma_{\rm{M}}} N\left(\gamma_{\rm{e}}\right)\rm{d}\gamma_{\rm{e}}$=$4\gamma n\left(r\right)$ and $\int_{\gamma_{\rm{m}}}^{\gamma_{\rm{M}}} N\left(\gamma_{\rm{e}}\right) \gamma_{\rm{e}} m_{\rm{e}} c^2\rm{d}\gamma_{\rm{e}}$=$4 \gamma\left(\gamma-1\right) n\left(r\right) m_{\rm{p}}c^2 \epsilon_{\rm_{e}}$, from which one obtains the minimum Lorentz factor,
\begin{equation}
\gamma_{\rm{m}}=\left(f_{\rm{p}} \frac{m_{\rm{p}}}{m_{\rm{e}}}\frac{\epsilon_{\rm{e}}}{\xi^{2-p_1}} \right)^{\frac{1}{p_1-1}} \beta^{-\frac{q\left(2-p_1\right)}{p_1-1}} \left(\gamma-1\right)^{\frac{1}{p_1-1}} \gamma^{-\frac{q\left(2-p_1\right)}{p_1-1}},
\end{equation}
where $f_{\rm{p}}=\frac{\left(2-p_1\right)\left(p_2-2\right)}{\left(p_1-1\right)\left(p_2-p_1\right)}$.

We calculate the break frequencies of synchrotron spectra $\nu_{\rm{m}}$, $\nu_{\rm{b}}$, $\nu_{\rm{c}}$  and the peak flux $F_{\nu,\rm{max}}$ according to the formula given by \citet{Wijers99}:
\begin{eqnarray}\label{vm1}
\nu_{\rm{m}} &=& \frac{x_{\rm{p}}}{1+z}\frac{q_{\rm{e}}B'}{\pi m_{\rm{e}}c} \gamma \gamma_{\rm{m}}^2,\\ \label{vbc}
\nu_{\rm{b,c}} &=& \frac{0.286}{1+z}\frac{q_{\rm{e}}B'}{\pi m_{\rm{e}}c} \gamma \gamma_{\rm{b,c}}^2,\\ \label{fmax}
F_{\nu,\rm{max}} &=& \frac{\phi_{\rm{p}} q_{\rm{e}}^3 \left(1+z\right)}{\sqrt{3} m_{\rm{e}} c^2 d_{\rm{L}}^2} B' \gamma r^3 n,
\end{eqnarray}
where $q_{\rm{e}}$ is the electron charge, $B'=\left(32\pi n m_{\rm{p}} c^2 \epsilon_{\rm{B}}\right)^{1/2}\gamma$ is the post-shock magnetic field density, $d_{\rm{L}}$ is the luminosity distance corresponding to the redshift $z$, $\gamma_{\rm{c}}=6\pi m_{\rm{e}}c/\left(\sigma_{\rm{T}}\gamma B'^2 t\right)$ is the cooling Lorentz factor of electrons. $x_{\rm{p}}$ and $\phi_{\rm{p}}$ are dimensionless peak frequency and peak flux, respectively. Their dependence on $p$ can be obtained from \citet{Wijers99}.

The calculation of break frequencies and peak flux given above depends on the hydrodynamic evolution of the shock. As the achromatic break at $t_{\rm{bk}}$ in the light curves of GRB 091127 is suggestive of a jet break, it is necessary to analyze its physical origin at first, which determines our treatment of the hydrodynamic evolution. A jet-break-like steepening in the light curves could be due to two effects. The first is the jet-edge effect happening at $\gamma\sim 1/\theta_{\rm{j}}$, after which the light curves have a steepening by $t^{-3/4}$ (or $t^{-1/2}$) for an ISM (or wind) medium \citep{Mes99}. This effect does not change the hydrodynamic evolution. The second effect is caused by sideways expansion, which has important effects on the hydrodynamics when $\gamma<1/\theta_{\rm{j}}$ is satisfied. The flux after the jet break decays as $t^{-p}$ for a normal electron spectrum with index $p>2$ \citep{Rhoads99,Sari99}. For a hard electron spectrum($1<p<2$), however, the decay slope after this kind of jet break is somewhat different \citep{Dai2001,Wang12}.

Nevertheless, numerical simulations and more sophisticated analytical treatments suggest that the sideways expansion of a relativistic jet is unimportant until $\gamma$ drops below $\sim 2$ \citep{Huang00,Granot01,Kumar03,Cann04,ZhangW09,decolle12,van12,Granot12}. Moreover, using the expression (their Equation 11) given by \citet{Wang12} who considered the effect of sideways expansion, the predicted decay slope of the light curves of GRB 091127 after $t_{\rm{bk}}$ is $\sim -2.4$ (when the parameters given in Section \ref{fitting} were used), which is much steeper than the observed. Therefore, we neglect the effect of sideways expansion on the hydrodynamic evolution in the relativistic phase.

Using the light curve decay indices given by \citet{Resmi08}, a wind-like circumburst medium can be easily excluded. So here we only consider the ISM case.
For simplicity, we consider the self-similar evolution of a spherical blast wave in the adiabatic case \citep{BM76}. The radius $r$ and bulk Lorentz factor $\gamma$ evolve as $r\left(t\right)=\left[17Et/4\pi n m_{\rm{p}}c\left(1+z\right)\right]^{1/4}$, and $\gamma\left(t\right)=\left[17E\left(1+z\right)/1024\pi n m_{\rm{p}} c^5 t^3\right]^{1/8}$ \citep{Sari98}, where $E$ is the isotropic equivalent energy of the jet, and $t$ is the time in the observer frame.  By substituting these expressions in Equations (\ref{vm1}),(\ref{vbc}) and (\ref{fmax}), one derives
\begin{eqnarray}
\nu_{\rm{m}} &=& 8.2\times 10^{6}\left(1833 f_{\rm{p}}\right)^{\frac{2}{p_1-1}} \left(37.2\right)^{\frac{1-q\left(2-p_1\right)}{p_1-1}} \frac{x_{\rm{p}}}{1+z}\xi^{\frac{-2\left(2-p_1\right)}{p_1-1}}\epsilon_{\rm{e}}^{\frac{2}{p_1-1}} \epsilon_{\rm{B,-2}}^{1/2}   \nonumber\\
             &  & E_{52}^{\frac{p_1-q\left(2-p_1\right)}{4\left(p_1-1\right)}} n_{0}^{\frac{p_1-2+q\left(2-p_1\right)}{4}} \left(\frac{t_{\rm{d}}}{1+z}\right)^{\frac{-3\left[p_1-q\left(2-p_1\right)\right]}{4\left(p_1-1\right)}}~~~~ \rm{Hz},  \\
\nu_{\rm{c}} &=& 1.5\times 10^{15} \epsilon_{\rm{B,-2}}^{-3/2} E_{52}^{-1/2} n_0^{-1} \left[\frac{t_{\rm{d}}}{1+z}\right]^{-1/2} \rm{Hz}, \\ \label{vb}
\nu_{\rm{b}} &=& 3.8\times 10^{5} ~\frac{\left(6.1\right)^{1+2q}}{1+z} \xi^2 \epsilon_{\rm{B,-2}}^{1/2} E_{52}^{\frac{1+q}{4}} n_{0}^{\frac{1-q}{4}} \left(\frac{t_{\rm{d}}}{1+z}\right)^{-\frac{3\left(1+q\right)}{4}} \rm{Hz}, \\
F_{\nu,\rm{max}} &=& 6.8\times 10^3 \phi_{\rm{p}} \epsilon_{\rm{B,-2}}^{1/2} E_{52} n_0^{1/2} d_{\rm{L,28}}^{-2} \left(1+z\right) ~~~\rm{\mu Jy},
\end{eqnarray}
where $t_{\rm{d}}$ is the time in days. The coefficients in above equations are somewhat different from those of \citet{Resmi08}, since they considered the evolution of $\theta_{\rm{j}}$ due to the lateral expansion of the jet, though it is not important in the ultra-relativistic phase as discussed above.

The evolution of the synchrotron flux density at a given frequency ($F_\nu$) relies on the order of the three break frequencies and which regime $\nu$ is in. For GRB091127, the SED analysis requires $\nu_{\rm{b}}$ to evolve from $\nu_{\rm{m}}<\nu_{\rm{opt}}<\nu_{\rm{b}}<\nu_{\rm{X}}<\nu_{\rm{c}}$ to $\nu_{\rm{m}}<\nu_{\rm{b}}<\nu_{\rm{opt}}<\nu_{\rm{X}}<\nu_{\rm{c}}$ between $\sim3$ ks and $\sim 110$ ks. Therefore, we just derive the flux density evolution in these regimes.
For $\nu_{\rm{m}}<\nu<\nu_{\rm{b}}<\nu_{\rm{c}}$,
\begin{equation}
F_{\nu}=F_{\nu,\rm{max}}\left(\frac{\nu}{\nu_{\rm{m}}}\right)^{-\frac{p_1-1}{2}}\propto t^{-\frac{3}{8}\left(p_1+p_1 q-2q\right)}. \label{fv1}
\end{equation}
For $\nu_{\rm{m}}<\nu_{\rm{b}}<\nu<\nu_{\rm{c}}$,
\begin{equation}
F_{\nu}=F_{\nu,\rm{max}}\left(\frac{\nu_{\rm{b}}}{\nu_{\rm{m}}}\right)^{-\frac{p_1-1}{2}} \left(\frac{\nu}{\nu_{\rm{b}}}\right)^{-\frac{p_2-1}{2}} \propto t^{-\frac{3}{8}\left(p_2+p_2 q-2q\right)}. \label{fv2}
\end{equation}

We don't consider the flux evolution in the non-relativistic phase, as the light curves of GRB 091127 show no evidence of further steepening at later times,
and we will show  in Section \ref{fitting} that the jet is still in the mildly relativistic phase at the end of the X-ray observations.

\section{PARAMETER CONSTRAINT AND LIGHT CURVE FITTING} \label{fitting}
Before constraining the free parameters ($\epsilon_{\rm{e}}$, $\epsilon_{\rm{B}}$, $\xi$, $E$ and $n$), we summarize the decay slopes and spectral indices of the afterglow of GRB 091127. For the X-ray afterglow, the decay slopes $\alpha_{\rm{X},1}=1.02\pm0.04$, $\alpha_{\rm{X},2}=1.61\pm0.04$, the spectral index  $\beta_{\rm{X}}=0.75\pm0.004$. While for the optical afterglow (we don't use the fitting results in the NIR bands), the average decay slopes $\alpha_{\rm{opt},1}=0.38\pm 0.04$, $\alpha_{\rm{opt},2}\sim1.6$, and the spectral index of $\beta_{\rm{opt}}=\beta_{\rm{X}}-0.5=0.25\pm0.004$ is used according to the SED analysis. Both X-ray and optical/NIR light curves have an achromatic break at $t_{\rm{bk}}\approx33$ ks. The spectral break frequency evolves as $\propto t^{-1.23\pm0.06}$.

At early times, $\nu_{\rm{m}}<\nu_{\rm{opt}}<\nu_{\rm{b}}<\nu_{\rm{X}}<\nu_{\rm{c}}$ is required, then the spectral indices of the electron energy distribution are given by $p_1=2\beta_{\rm{opt}}+1=1.5\pm0.01$, and $p_2=2\beta_{\rm{X}}+1=2.5\pm0.01$.  With these values and according to Equations (\ref{vb}), (\ref{fv1}) and (\ref{fv2}), the decay indices of $\nu_{\rm{b}}$, $F_{\nu_{\rm{opt}}}$ and $F_{\nu_{\rm{X}}}$ are only functions of $q$. Therefore, the value of $q$ is overconstrained. As the initial decay slope of optical afterglow was not well fitted \citep{Filgas11}, we use the decay indices of $\nu_{\rm{b}}$ and $F_{\nu_{\rm{X}}}$ to constrain $q$ and give a consistency check using the optical data. With Equations (\ref{vb}) and (\ref{fv2}) and the observed decay indices, we get
\begin{eqnarray}\label{q1}
\frac{3\left(1+q\right)}{4} & = & 1.23 \pm 0.06,  \\  \label{q2}
\frac{3}{8}\left(p_2+p_2 q-2q\right) & = & 1.02\pm0.04.
\end{eqnarray}
From Equations (\ref{q1}) and (\ref{q2}), we obtain $q=0.64\pm0.08$ and $q=0.44\pm0.21$, respectively. However, the former is preferred, since the left hand side of Equation (\ref{q1}) is much more dependent on $q$ than that of Equation (\ref{q2}). With this value of $q$ and according to Equations (\ref{fv1}) and(\ref{fv2}), the derived decay indices are $\alpha_{\rm{opt},1}=0.44\pm0.02$ and $\alpha_{\rm{X},1}=1.06\pm0.02$, which are consistent with the observed values within $1\sigma$ errors. Therefore, we adopt $q=0.64$ in the following calculations.

The synchrotron flux in the optical bands is given by
\begin{equation}
F_{\nu_{\rm{opt}}}=F_{\nu,\rm{max}} \left(\frac{\nu_{\rm{opt}}}{\nu_{\rm{m}}}\right)^{-\beta_{\rm{opt}}}=263.9 ~\xi_4^{-1/2} \epsilon_{\rm{e},-1} \epsilon_{\rm{B},-2}^{5/8} E_{52}^{1.15} n_0^{0.48} t_{\rm{d}}^{-0.44} \left(\frac{\nu_{\rm{opt}}}{\nu_{\rm{r}}}\right)^{-\beta_{\rm{opt}}} ~~\rm{\mu Jy}. \label{fopt}
\end{equation}
By requiring the r-band flux be $827 ~\mu\rm{Jy}$ at 4000 s, we obtain
\begin{equation}
\xi_4^{-1/2} \epsilon_{\rm{e},-1} \epsilon_{\rm{B},-2}^{5/8} E_{52}^{1.15} n_0^{0.48} = 0.8. \label{pa_fr}
\end{equation}
According to the SED analysis, $\nu_{\rm_{b}}$ should be $\sim28.7$ eV at 3404 s \citep{Filgas11}, then with Equation (9), we have
\begin{equation}
\xi_4^2 \epsilon_{\rm{B},-2}^{1/2} E_{52}^{-0.41} n_0^{0.1} =0.05. \label{pa_vb}
\end{equation}
Finally, $\nu_{\rm{c}}$ should have not crossed the X-band at the last measurement of the X-ray afterglow, i.e. $\nu_{\rm{c}}\left(4\times10^6 ~\rm{s}\right)> 10 ~\rm{keV}$. With Equation (8), we get
\begin{equation}
\epsilon_{\rm{B},-2}^{-3/2} E_{52}^{-1/2} n_0^{-1} > 1.4 \times 10^4. \label{pa_vc}
\end{equation}
From Equations (\ref{pa_fr}), (\ref{pa_vb}) and (\ref{pa_vc}), one derives
\begin{eqnarray} \label{cond1}
\epsilon_{\rm{B},-2} n_{0}^{2/3} & = & 0.28 \epsilon_{\rm{e},-1}^{-4/3} E_{52}^{-1.4}, \\  \label{cond2}
\xi_4 & =& 0.22 ~\epsilon_{\rm{B},-2}^{-1/4} E_{52}^{0.21} n_0^{-0.05}, \\ \label{cond3}
\epsilon_{\rm{e},-1} & > & 45.6 E_{52}^{-0.8}.
\end{eqnarray}
By requiring $\epsilon_{\rm{e}}<1$ and a not-too-low efficiency of the prompt radiation (we simply assume $\eta_\gamma=E_{\gamma}/\left(E_\gamma+E\right)>5\%$ ), we derive $6.6 < E_{52} <20.9$  and $\epsilon_{\rm{e},-1}>4.2$ from Equation (\ref{cond3}). Here we take $E_{52}=20$ and $\epsilon_{\rm{e},-1}=4.5$. By substituting these values in Equations (\ref{cond1}), we get $\epsilon_{\rm{B},-2} n_{0}^{2/3}=5.7\times10^{-4}$. The values of $\epsilon_{\rm{B}}$ and $n$ can not be well constrained, since both of them are highly uncertain parameters and vary over several orders of magnitude. Without loss of generality, we adopt $n_0=1$, then we obtain $\epsilon_{\rm{B},-2}=5.7\times10^{-4}$. By substituting these values in Equation (\ref{cond2}), we get $\xi_4=2.7$. The value of $\xi$ can be well constrained, it is around $2\times10^4$, varying within a factor of 2, since it is weakly dependent on other parameters according to Equation (\ref{cond2}).

We note that the value of $\epsilon_{\rm{B}}$ obtained above is much smaller than usually assumed ($\sim10^{-3}$--$10^{-2}$). However, such a small value may be more common according to the recent statistic results given by \citet{Santana14}. Using X-ray and optical afterglow observations, they found the distribution of $\epsilon_{\rm{B}}$ has a range of $\sim 10^{-8} - 10^{-3}$, with a median value $\sim$ few $\times 10^{-5}$. Another separate work using the radio data also support this result \citep{Duran14}.

 Since we interpret the achromatic break at $t_{\rm{bk}}$ as a jet break, we can estimate the half-opening angle of the jet according to $\theta_{\rm{j}}\sim \gamma\left(t_{\rm{bk}}\right)^{-1}$, thus we have
\begin{equation}
\theta_{\rm{j}}=9.4~ E_{52}^{-1/8} n_0^{1/8} \left(\frac{t_{\rm{bk,d}}}{1+z}\right)^{3/8} ~\rm{deg}=3.9 ~\rm{deg}.
\end{equation}
The bulk Lorentz factor at the end of the X-ray observations is $\gamma\left(4\times10^6 \rm{s}\right)=2.5$, which is still mildly relativistic. Therefore, our explanation of the afterglow of GRB091127 in the relativistic case is self-consistent.

As a whole, the DPLH model can roughly explain the main property of the spectral evolution in the optical/NIR bands of GRB 091127. However, the analytic treatment of the flux evolution given by Equations (\ref{fv1}) and (\ref{fv2}) is still too simple to describe the detailed spectral evolution and the smooth break at around $t_{\rm{bk}}$ in the optical/NIR light curves, which requires a very smooth spectral break at $\nu_{\rm{b}}$. In fact,
when the equal arrival time surface effect (ETS) of the relativistic ejecta is considered \citep{Waxm97,Sari1998,Panai98}, the spectral and temporal breaks are rather smooth \citep{Granot99,Huang07}, and can be described by a smoothly broken power-law at the spectral break \citep{Granot02}. Besides, for a uniform jet with a
sharp cutoff at the edge, the jet break caused by the edge effect is very sharp as well. For GRB 091127, the predicted decay slope after the jet break is
$\sim 1.8$, which is still too steep compared with the observed value ($\sim1.6$). However, the more realistic jet may be structured \citep{Rossi02,Zhang02,Granot03,Pescalli15,Salafia15}, which could smoothen the jet break.

Therefore, instead of using Equations (\ref{fv1}) and (\ref{fv2}), we fit the multi-band light curves with the following expression
\begin{equation}
F_{\nu}=F_0 \left[\left(\frac{\nu}{\nu_{\rm{b}}}\right)^{s_1 \beta_{\rm{opt}}}+\left(\frac{\nu}{\nu_{\rm{b}}}\right)^{s_1 \beta_{\rm{X}}}\right]^{-\frac{1}{s_1}}
\left[1+\left(\frac{t}{t_{\rm{bk}}}\right)^{s_2\Delta\alpha}\right]^{-\frac{1}{s_2}},  \label{fitfunction}
\end{equation}
 where the first smoothly broken power-law describes the spectral evolution with a smooth break $\nu_{\rm{b}}$, while the second describes the smooth jet break caused by jet-edge effect.
$F_0$ is the normalization, which can be obtained by requiring $F_{\nu_{\rm{r}}}\left(4000\rm{s}\right)=827 \mu \rm{Jy}$. $s_1$ and $s_2$ are smoothness parameters, we take $s_1=2.2$ according to the SED fitting \citep{Filgas11}, and take $s_2=2$
to describe the sharpness of the X-ray light curve at $t_{\rm{bk}}$. The parameter $\Delta\alpha=3/4$ accounts for the slope difference before and after the jet break. Here we still use $\nu_{\rm{b}}$ given by Equation (\ref{vb}) with the parameters obtained above and neglect the small correction due to the ETS effect.

\begin{figure}[htbp]
\begin{center}
\includegraphics[width=15cm]{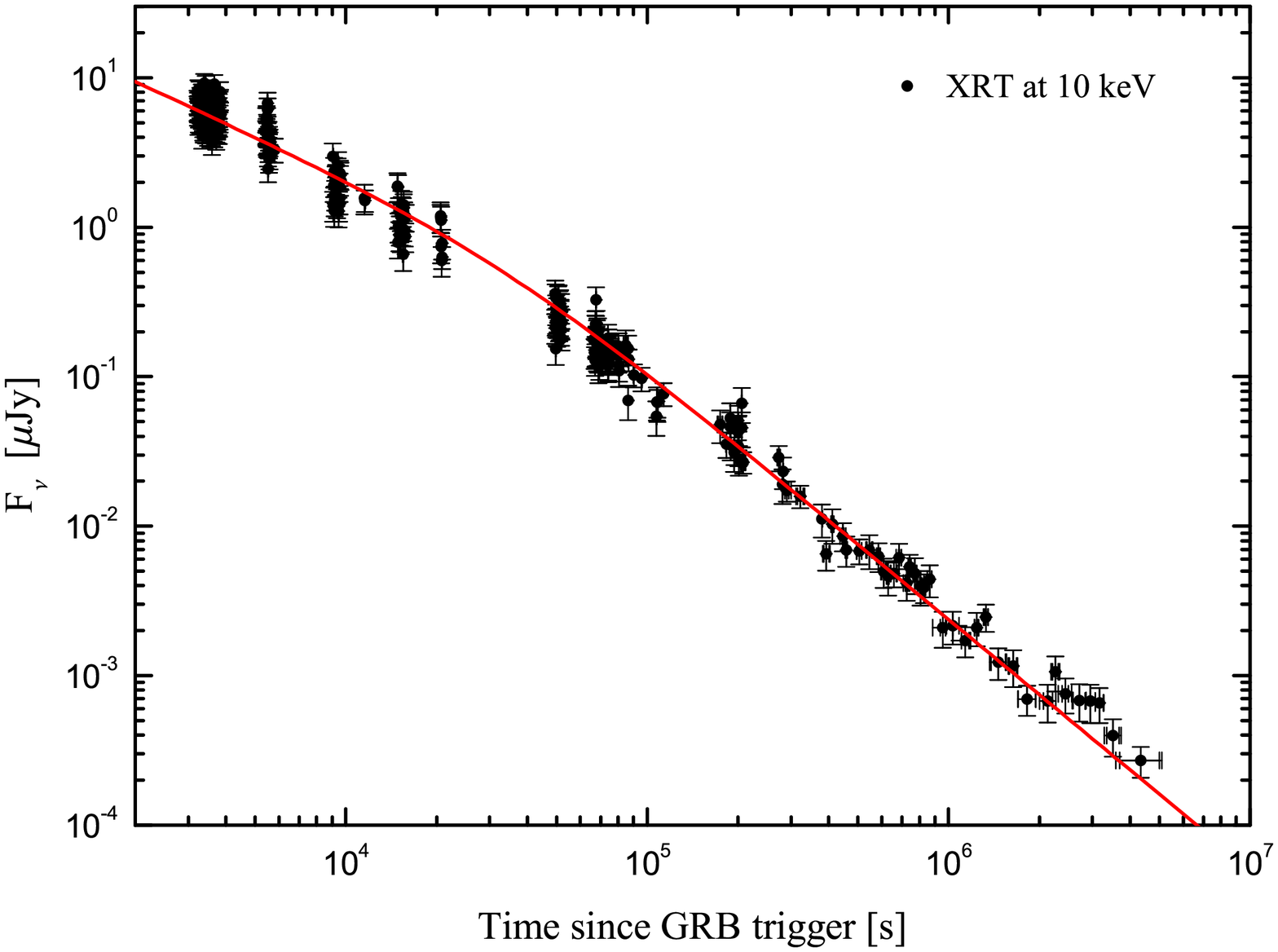}
\caption{Fit of the X-ray light curve at 10 keV. The data (filled circles) are taken from http://www.swift.ac.uk/xrt\_curves/00377179/ \citep{Evans09a}. The solid line is the theoretical light curve given by Equation (\ref{fitfunction}).
 \label{XRTfitting}}
\end{center}.
\end{figure}

\begin{figure}[htbp]
\begin{center}
\includegraphics[width=15cm]{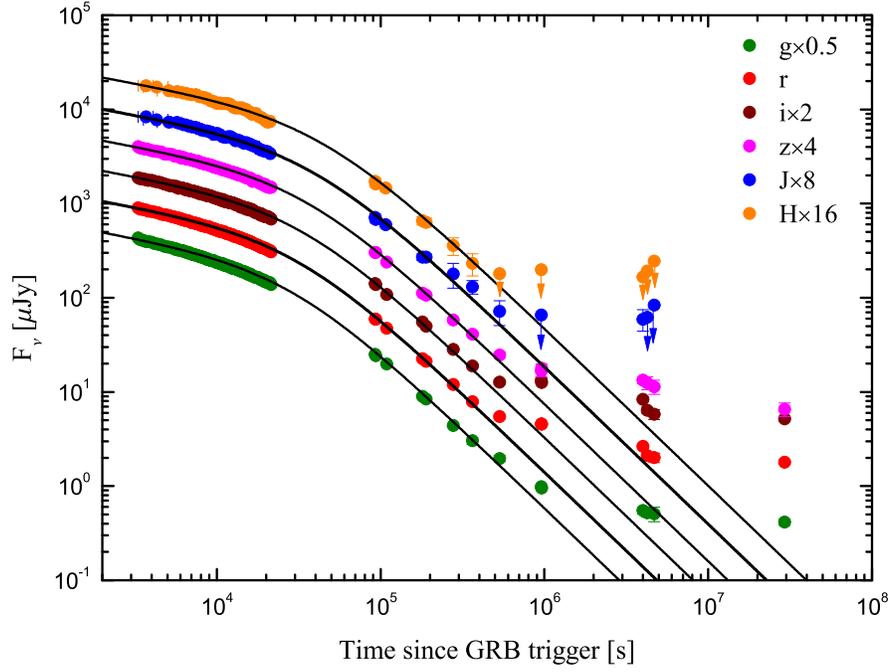}
\caption{Fit of the multi-band optical/NIR light curves observed by GROND.  The observational data (filled circles) are taken from \citet{Filgas11}. Adjacent light curves have been offset by a factor of 2 for clarity with r band unscaled. The solid lines are theoretical light curves given by Equation (\ref{fitfunction}).  \label{optfitting}}
\end{center}
\end{figure}

As can be seen from Figure \ref{XRTfitting} and Figure \ref{optfitting}, except for the SN components we are not concerned here, the multi-band afterglow light curves of GRB 091127 can be well fitted by our theoretical motivated equation of (\ref{fitfunction}). This is to be expected, though. Since the first smoothly broken power-law in this fitting function is just the one used by \citet{Filgas11} in their SED analysis.  However, we give a physical meaning of the observed spectral break in this paper, which is expressed by Equation (\ref{vb}) instead of a free parameter in their SED fitting.

\section{CONCLUSION AND DISCUSSION} \label{conclusion}
GRB 091127, with well-sampled broad-band afterglow data, shows evidence of a hard electron spectrum and strong spectral evolution, with a spectral break moving from high to lower energies. The spectral break evolves much faster than the cooling break even in the ISM case, which challenge the standard afterglow model. In this paper, using the DPLH model and an assumed evolution function of the injection break, we interpreted the observed spectral break as the injection break frequency. The observed spectral evolution is due to this injection break crossing the optical/NIR bands. In addition, we interpreted the achromatic break at around 33 ks as a jet break caused by the jet-edge effect. We have shown that the multi-band observational data can be satisfactorily fitted in our framework.

Our model is intrinsically different from that of \citet{Filgas11}, although both assumed a hard electron spectrum.  \citet{Filgas11} used the SPLH model, they interpreted the observed spectral break as the cooling break, but requiring $\epsilon_{\rm{B}}$ to evolve with time. While we used the DPLH model, and explained the observed spectral break as the injection break, without requiring evolving microphysical parameters. Currently, we have little knowledge of the injection break and its evolution, and in-depth numerical simulation studies on electron acceleration process may help to solve these issues.

\citet{Resmi08} used the same model to explain the afterglows of three pre-{\it Swift} GRBs. However, the parameters of GRB 091127 are much different from those given by \citet{Resmi08}. Firstly, the optical spectral indices of their sample are in the range of $0.6-0.9$. To be modeled with a hard electron spectrum, all the cooling frequencies $\nu_{\rm{c}}$ of their sample were assumed to be below the optical band at the beginning of observations. In our study, the cooling break of GRB 091127 was required to be above the X-ray band even at late times. Secondly, all their injection break frequencies $\nu_{\rm{b}}$ were assumed to be above the X-ray band, while for GRB 091127, $\nu_{\rm{b}}$ was required to be between the optical and X-ray band at early times and to cross the optical band at later times. Thirdly, since $\nu_{\rm{c}}\propto \epsilon_{\rm{B}}^{-3/2}$ and $\nu_{\rm{b}}\propto \epsilon_{\rm{B}}^{1/2}$, the relatively low $\nu_{\rm{c}}$ and high $\nu_{\rm{b}}$ determine that their derived $\epsilon_{\rm{B}}$ ($\sim 0.01-0.2$) are much lager than ours ($\sim 10^{-6}-10^{-5}$). Finally, all their values of $q$ are lager than 1, while for GRB 091127, $q$ is smaller ($\sim0.6$). Due to these differences among parameters, the afterglow of GRB 091127 reveals much different properties from those in \citet{Resmi08}. In this case, we see the evolution of the injection break for the first time.

Besides GRB 091127, the other two GRB afterglows, GRB 060908 \citep{Covino10} and GRB 140515A \citep{Mela15}, observed by the {\it Swift} satellite also show very flat spectra in the optical band ($\beta_{\rm{opt}}\sim0.3$), and could be explained with a hard electron spectrum \citep{Wang12,Mela15}. Such GRBs, however, are still lacking. More observations of GRB afterglows with a hard electron spectrum and further developments in the area of simulations of Fermi acceleration process in relativistic shocks will help us understand the origin of the hard electron distribution.

\acknowledgments

 We acknowledge the anonymous referee for his/her helpful comments and suggestions. We thank Tan Lu for encouraging support. Our work made use of data supplied by the UK Swift Science Data Centre at the University of Leicester. This study was supported by the National Basic Research Program of China with Grant No. 2014CB845800,  and by the National Natural Science Foundation of China with Grant No. 11473012, No. 11475085 and No. 11275097.

\clearpage

\begin{thebibliography}{}
\bibitem[Achterberg et al.(2001)]{Acht01} Achterberg, A., Gallant, Y. A., Kirk, J. G., \& Guthmann, A. W. 2001, \mnras, 328, 393
\bibitem[Band et al.(1993)]{Band93} Band, D., Matteson, J., Ford, L., et al. 1993, \apj, 413, 281
\bibitem[Baring(2004)]{Bar04} Baring, M. G. 2004, NuPhS, 136, 198
\bibitem[Barniol Duran(2014)]{Duran14} Barniol Duran, R. 2014, \mnras, 442, 3147
\bibitem[Bednarz et al.(1998)]{Bed98} Bednarz, J., \& Ostrowski, M. 1998, PhRvL, 80, 3911
\bibitem[Beuermann et al.(1999)]{Beue99} Beuermann, K., Hessman, F. V., Reinsch, K., et al. 1999, \aap, 352, L26
\bibitem[Bhattacharya(2001)]{B01} Bhattacharya D. 2001, BASI, 29, 107
\bibitem[Bhattacharya \& Resmi(2004)]{B04} Bhattacharya D., Resmi L. 2004, in ASP Conf. Ser. 312, Evolution of an
Afterglow with a Hard Electron Spectrum, ed. M. Feroci et al. (San Francisco, CA: ASP), 411
\bibitem[Bj\"ornsson et al.(2002)]{Bjor02} Bj\"ornsson G., Hjorth J., Pedersen K., \& Fynbo J. U. 2002, ApJL, 579, L59
\bibitem[Blandford \& McKee(1976)]{BM76} Blandford, R. D., \& McKee, C. F. 1976, PhFl, 19, 1130
\bibitem[Bulter et al.(2003)]{But03} Butler N. R., Marshall H. L., Ricker G. R., et al. 2003, \apj, 597, 1010
\bibitem[Cannizzo et al.(2004)]{Cann04} Cannizzo, J. K., Gehrel, N., \& Vishniac, E. T. 2004, \apj, 601, 380
\bibitem[Chevalier \& Li(2000)]{Cheva2000} Chevalier, R. A., \& Li, Z.-Y. 2000, \apj, 536, 195
\bibitem[Cobb et al.(2010)]{Cobb10} Cobb, B. E., Bloom, J. S., Perley, D. A., et al. 2010, \apj, 711, 641
\bibitem[Covino et al.(2010)]{Covino10} Covino S., Campana S., Conciatore M. L., et al. 2010, \aap, 521, A53
\bibitem[Covino et al.(2003)]{Cov03} Covino, S., Malesani, D., Tavecchio, F., et al. 2003, \aap, 404, L5
\bibitem[Cowsik et al.(2001)]{Cow01} Cowsik, R., Prabhu, T. P., Anupama, G. C., et al. 2001, BASI, 29, 157
\bibitem[Cucchiara et al.(2009)]{Cucch09} Cucchiara, A., Fox, D., Levan, A., et al. 2009, GCN, 10202, 1
\bibitem[Curran et al.(2010)]{Curran10} Curran, P. A., Evans, P. A., de Pasquale, M., Page, M. J., \& van der Horst, A. J. 2010, ApJL, 716, L135
\bibitem[Curran et al.(2009)]{Curran09} Curran, P. A., Starling, R. L. C., van der Horst, A. J., \& Wijers, R. A. M. J. 2009, \mnras, 395, 580
\bibitem[Dai \& Cheng(2001)]{Dai2001} Dai, Z. G., \& Cheng, K. S. 2001, ApJL, 558, L109
\bibitem[Dai \& Lu(1998)]{Dai98} Dai, Z. G., \& Lu, T. 1998, \aap, 333, L87
\bibitem[de Colle et al.(2012)]{decolle12} de Colle, F., Ramirez-Ruiz, E., Granot, J., \& Lopez-Camara, D. 2012, \apj, 751, 57
\bibitem[Evans et al.(2009a)]{Evans09a} Evans, P. A., Beardmore, A. P., Page, K. L., et al. 2009a, \mnras, 397, 1177
\bibitem[Evans et al.(2009b)]{Evans09b} Evans, P. A., Page, K. L., \& Troja, E. 2009b, GCN, 10201, 1
\bibitem[Fermi(1954)]{Fermi54} Fermi, E. 1954, \apj, 119, 1
\bibitem[Filgas et al.(2011)]{Filgas11} Filgas, R., Greiner, J., Schady, P., et al. 2011, \aap, 537, A57
\bibitem[Granot \& Kumar(2003)]{Granot03} Granot, J., \& Kumar, P. 2003, \apj, 591, 1086
\bibitem[Granot et al.(2001)]{Granot01} Granot, J., Miller, M., Piran,  T., Suen, W. M., \& Hughes, P. A. 2001,in Era
ESO Afterglow Symp., Gamma-ray Bursts in the Afterglow, Relativistic Jet, ed. E. Costa, F. Frontera \& J. Hjorth (Berlin: Springer), 312
\bibitem[Granot et al.(1999)]{Granot99} Granot, J., Piran, T., \& Sari, R. 1999, \apj, 513, 679
\bibitem[Granot \& Piran(2012)]{Granot12} Granot, J., \& Piran, T. 2012, \mnras, 421, 570
\bibitem[Granot \& Sari(2002)]{Granot02} Granot, J., \& Sari, R. 2002, \apj, 568, 820
\bibitem[Huang et al.(2006)]{Huang06} Huang, Y. F., Cheng, K. S., \& Gao, T. T. 2006, \apj, 637, 873
\bibitem[Huang et al.(2000)]{Huang00} Huang, Y. F., Gou, L. J., Dai, Z. G., Lu, T. 2000, \apj, 543, 90
\bibitem[Huang et al.(2007)]{Huang07} Huang, Y. F., Lu, Y., Wong, A. Y. L., \& Cheng, K. S. 2007, ChJAA, 7, 397
\bibitem[Ioka et al.(2006)]{Ioka06} Ioka K., Toma K., Yamazaki R., \& Nakamura N. 2006, \aap, 458, 7
\bibitem[Kirk et al.(2000)]{Kirk00} Kirk, J. G., Guthmann, A. W., Gallant, Y. A., \& Achterberg, A. 2000, \apj, 542, 235
\bibitem[Kong et al.(2010)]{Kong10} Kong, S. W., Wong, A. Y. L., Huang, Y. F., Cheng, K. S. 2010, \mnras, 402, 409
\bibitem[Kumar \& Granot(2003)]{Kumar03} Kumar, P., \& Granot, J. 2003, \apj, 591, 1075
\bibitem[Kumar \& Zhang(2015)]{Kumar15} Kumar, P., \& Zhang, B. 2015, Physics Reports, 561, 1
\bibitem[Lemoine \& Pelletier(2003)]{Lem03} Lemoine, M., \& Pelletier, G. 2003, ApJL, 589, L73
\bibitem[Masetti et al.(2001)]{Mas01} Masetti, N., Palazzi, E., Pian, E., et al. 2001, \aap, 374, 382
\bibitem[Melandri et al.(2015)]{Mela15} Melandri, A, Bernardini, M. G., D'Avanzo, P., et al. 2015, arxiv: 1506.03079
\bibitem[M\'esz\'aros \& Rees(1993)]{Mes93} M\'esz\'aros, P., \& Rees, M. J. 1993, \apj, 405, 278
\bibitem[M\'esz\'aros \& Rees(1997)]{Mes97} M\'esz\'aros, P., \& Rees, M. J. 1997, \apj, 476, 232
\bibitem[M\'esz\'aros \& Rees(1999)]{Mes99} M\'esz\'aros, P., \& Rees, M. J. 1999, \mnras, 306, L39
\bibitem[Misra et al.(2005)]{Misra05} Misra K., Resmi L., Pandey S. B., Bhattacharya D., \& Sagar R. 2005, BASI, 33, 487
\bibitem[Nousek et al.(2006)]{Nou06} Nousek, J. A., Kouveliotou, C., Grupe, D., et al. 2006, \apj, 642, 389
\bibitem[Panaitescu \& Kumar(2001a)]{Panai01a} Panaitescu, A., \& Kumar P. 2001a, ApJL, 560, L49
\bibitem[Panaitescu \& Kumar(2001b)]{Panai01} Panaitescu, A., \& Kumar P. 2001b, \apj, 554, 667
\bibitem[Panaitescu \& Kumar(2002)]{Panai02} Panaitescu, A., \& Kumar, P. 2002, ApJ, 571, 779
\bibitem[Panitescu \& M\'esz\'aros(1998)]{Panai98} Panaitescu, A., \& M\'esz\'aros, P. 1998, ApJL, 493, L31
\bibitem[Panaitescu et al.(2006)]{Panai06} Panaitescu, A., M\'esz\'aros, P., Burrows, D., et al. 2006, \mnras, 369, 2059
\bibitem[Pescalli et al.(2015)]{Pescalli15} Pescalli, A., Ghirlanda, G., Salafia, O., et al. 2015, \mnras, 447, 1911
\bibitem[Rees \& M\'esz\'aros(1992)]{Rees92} Rees, M. J., \& M\'esz\'aros, P. 1992, \mnras, 258, 41P
\bibitem[Ress \& M\'esz\'aros(1998)]{Rees98} Rees, M. J., \& M\'esz\'aros, P. 1998, ApJL, 496, L1
\bibitem[Resmi \& Bhattacharya(2008)]{Resmi08} Resmi, L., \& Bhattacharya, D. 2008, \mnras, 388, 144
\bibitem[Rhoads(1999)]{Rhoads99} Rhoads, J. E. 1999, \apj, 525, 737
\bibitem[Rossi et al.(2002)]{Rossi02} Rossi, E., Lazzati, D., \& Rees, M. J. 2002, \mnras, 332, 945
\bibitem[Sagar et al.(2001)]{Sag01} Sagar, R., Stalin, C. S., Bhattacharya, D., et al. 2001, BASI, 29, 91
\bibitem[Salafia et al.(2015)]{Salafia15} Salafia, O. S., Ghisellini, G., Pescalli, A., Ghirlanda, G., \& Nappo, F. 2015, \mnras, 450. 3549
\bibitem[Santana et al.(2014)]{Santana14} Santana, R., Barniol Duran, R., \& Kumar, P. 2014, \apj, 785, 29
\bibitem[Sari(1998)]{Sari1998} Sari, R. 1998, ApJL, 494, L49
\bibitem[Sari \& M\'esz\'aros(2000)]{Sari00} Sari, R., \& Me\'sza\'ros, P. 2000, ApJL, 535, L33
\bibitem[Sari et al.(1998)]{Sari98} Sari, R., Piran, T., \& Narayan, R. 1998, ApJL, 497, L17
\bibitem[Sari et al.(1999)]{Sari99} Sari, R., Piran, T., \& Halpern, J. P. 1999, ApJL, 519, L17
\bibitem[Shen et al.(2006)]{Shen06} Shen, R., Kumar, P., \& Robinson, E. L. 2006, \mnras, 371, 1441
\bibitem[Smith et al.(2009)]{Smith09} Smith, R. J., Kobayashi, S., Guidorzi, C., \& Mundell, C. G. 2009, GCN, 10192, 1
\bibitem[Spitkovsky(2008)]{Spit08} Spitkovsky, A. 2008, ApJL, 682, L5
\bibitem[Stamatikos et al.(2009)]{Stama09} Stamatikos, M., Barthelmy, S. D., Baumgartner, W. H., et al. 2009, GCN, 10197, 1
\bibitem[Stanek et al.(2001)]{Stan01} Stanek, K. Z., Garnavich, P. M., Jha, S., et al. 2001, \apj, 563, 592
\bibitem[Starling et al.(2008)]{Starl08} Starling, R. L. C., Van der Horst, A. J., Rol, E., et al. 2008, \apj, 672, 433
\bibitem[Th\"one et al.(2009)]{Thone09} Th\"one, C. C., Goldoni, P., Covino, S., et al. 2009, GCN, 10233, 1
\bibitem[Troja et al.(2009)]{Troja09} Troja, E., Barthelmy, S. D., Baumgartner, W. H., et al. 2009, GCN, 10191, 1
\bibitem[Troja et al.(2012)]{Troja12} Troja, E., Sakamoto, T., Guidorzi, C., et al. 2012, \apj, 761, 50
\bibitem[Updike et al.(2009)]{Updike09} Updike, A., Rossi, A., Rau, A., et al. 2009, GCN, 10195, 1
\bibitem[van der Horst et al.(2014)]{Van14} van der Horst, A. J., Paragi, Z., de Bruyn, A. G., et al. 2014, \mnras, 444, 3151
\bibitem[van Eerten et al.(2012)]{van12} van Eerten, H., van der Horst, A., \& MacFadyen, A. 2012, \apj, 749, 44
\bibitem[Wang et al.(2012)]{Wang12} Wang, Y., Fan, Y.-Z., Wei, D.-M., \& Stefano, C. 2012, ChA\&A, 36, 148
\bibitem[Waxman(1997)]{Waxm97} Waxman, E. 1997, ApJL, 491, L19
\bibitem[Wijers \& Galama(1999)]{Wijers99} Wijers R. A. M. J., \& Galama T. J. 1999, \apj, 523, 177
\bibitem[Zhang et al.(2006)]{Zhang06} Zhang, B., Fan, Y. Z., Dyks, J., et al. 2006, \apj, 642, 354
\bibitem[Zhang \& M\'esz\'aros(2001)]{Zhang01} Zhang, B., \& M\'esz\'aros, P. 2001, ApJL, 552, L35
\bibitem[Zhang \& M\'esz\'aros(2002)]{Zhang02} Zhang, B., \& M\'esz\'aros, P. 2002, \apj, 571, 876
\bibitem[Zhang \& MacFadyen(2009)]{ZhangW09} Zhang, W., \& MacFadyen, A. 2009, \apj, 698, 1261
\end{thebibliography}
\end{document}